On the Way to More Convenient Description of Drug-Plasma Protein Binding

Piekarski S.,[1] Rewekant M.[2]

[1]Institut of Fundamental Technological Research Polish Academy of Science, Warsaw, Poland
[2]Medical University of Warsaw, Department of Pharmacodynamics, Warsaw, Poland

# Abstract

The theoretical case is considered where the total amount of plasma protein is conserved, but the drug is eliminated after its single application.

After a single drug application at time $t = 0$, the total drug concentration is measured at times $t_i,\ldots,t_k$ and the total drug concentration at time $t_i$ is denoted by $\varphi_i$.

Our discussion is limited to one protein binding site. The quantity of plasma protein ($\Lambda$), the association constant ($K_a$) and the total concentration of the drug $\varphi_i$ at time $t_i$ are considered as independent variables.

Free drug concentration, plasma protein bound concentration and free drug fraction are given as functions of these "new" variables. The aim of this communication is to derive the formula that allows to calculate the free drug concentration at any time after the drug application, based on 3 parameters: the association constant of the drug, the total plasma concentration of the drug and the concentration of the protein.

If the plasma protein quantity ($\Lambda$) and the association constant ($K_a$) are known, then from the knowledge of the total drug concentration $\varphi_i$ at time $t_i$ it is possible to determine the free drug concentration at time $t_i$.

**Keywords**: plasma protein binding; association constant; free drug concentration; unbound fraction

# Introduction

Pharmacological effect of a drug over time is the result of pharmacokinetic and pharmacodynamic processes determined by concentration of a free, non-protein bound form of the drug in the vascular bed. In experiments, the total concentration of a drug is usually determined as a sum of the free concentration of a drug ($c_{free}$) and plasma protein bound concentration ($c_{bound}$)(1).

The total drug concentration $\varphi$ is a sum of $c_{free}$ and $c_{bound}$. The ratio of $c_{free}$ and $\varphi$ is the unbound fraction and is denoted by $\varepsilon_{free}$.

Plasma protein bound fraction can also be defined as $\varepsilon_{bound} = 1 - \varepsilon_{free}$.

Information about the level of protein binding of the drug in serum is obtained from studies *in vitro* as parameters such as the association constant ($K_a$), the plasma protein binding (PPB) or the unbound fraction $f_{uP}$ (2).

Recently, a number of papers appeared which criticize application of the unbound fraction determined *in vitro* to PK parameters calculation in clinical practice (3).

In our work we have derived a formula that allows to calculate the free drug concentration at any time after the drug application, based on 3 parameters: the association constant of the drug, the total plasma concentration of the drug and the concentration of the protein. The model has been constructed under Goldberg-Waage law of mass action (4).

# Assumption and basic equations

In our discussion we assume a single intravenous dose application, determination of the total drug concentration in blood serum within the limit of detection in time intervals from $t_0$ to $t_i$, and determination of the total protein concentration. In the calculations we also use the association constant ($K_a$) of the drug. It`s inverse, the dissociation constant $K_d$ is a binding parameter (5).

Let us consider the case when the drug is eliminated, but the total amount of plasma protein is conserved (*in vivo*). The total amount of the protein ($\lambda$) is the sum of the protein bound with the drug ($c_{bound,i}$) and the free form of the protein ($e_i$).

$$c_{bound,i} + e_i = \Lambda = const. \qquad (1)$$

From (1) the following is obtained:

$$e_i = \Lambda - c_{bound,i} \qquad (2)$$

As is well known, the law of mass action has the form:

$$c_{bound,i} = K_a \times e_i \times c_{free,i} \quad (3)$$

After inserting (2) into (3) one obtains:

$$c_{bound,i} = K_a(\Lambda - c_{bound,i}) \times c_{free,i} \quad (4)$$

And therefore:

$$c_{bound,i} = \frac{K_a \times \Lambda \times c_{free,i}}{1 + K_a \times c_{free,i}} \quad (5)$$

## New approach

The sum of both forms of a drug g ($c_{bound,i}$ + $c_{free,i}$) must be equal to the total concentration of the drug, which will be denoted by $\varphi_i$:

$$\Phi_i = c_{bound,i} + c_{free,i} = \frac{K_a \times \Lambda \times c_{free,i}}{1 + K_a \times c_{free,i}} + c_{free,i} \quad (6)$$

Equation (6) implies the relation:

$$K_a \times c_{free,i}^2 + c_{free,i}(K_a\Lambda + 1 - K_a + 1 - K_a \times \Phi_i) = 0 \quad (7)$$

which is a trinomial square of $c_{free,i}$ with parameters $K_a$, $\Lambda$ and $\varphi_i$.

$$c_{free,i,P} = \frac{-[K_a \times (\Lambda - \Phi_i) + 1] + \sqrt{[K_a(\Lambda - \Phi_i) + 1]^2 + 4 \times K_a \times \Phi_i}}{2K_a} \quad (8)$$

$C_{bound,i}$ can be obtained by inserting (8) into (5). Afterwards, the percentage of free and bound with protein forms of the drug ($\varepsilon_{free}$ and $\varepsilon_{bound}$) can be calculated at any time after i.v. application of a single drug dose.

$$\varepsilon_{bound,i} = \frac{2K_a \times \Lambda}{1 + K_a \times \Lambda + K_a \times \Phi_i + \sqrt{[K_a \times (\Lambda - \Phi_i) + 1]^2 + 4 \times K_a \times \Phi_i}} \quad (9)$$

And for unbound fraction:

$$\varepsilon_{bound,i} = \frac{1 - K_a \times \Lambda + K_a \times \Phi_i + \sqrt{[K_a(\Lambda - \Phi_i) + 1]^2 + 4 \times K_a \times \Phi_i}}{1 + K_a \times \Lambda + K_a \times \Phi_i + \sqrt{[K_a(\Lambda - \Phi_i) + 1]^2 + 4 \times K_a \times \Phi_i}} \quad (10)$$

It should be emphasized that both bound fractions are now the functions of the variables $K_a$, $\Lambda$ and $\varphi_i$.

For small values of $\varphi_i$:

$$\varphi_i \approx 0 \tag{11}$$

the two last relations simplify to:

$$\varepsilon_{bound} = \frac{K_a \times \Lambda}{1 + K_a \times \Lambda} \tag{12},$$

$$\varepsilon_{free} = \frac{1}{1 + K_a \times \Lambda} \tag{13}$$

It is worth to note that assuming the (11) limitation the sum of (12) and (13) is equal to 1 which indirectly proves the correctness of the calculations.

## Disscussion

The proposed formula enables calculation of the free drug concentration and the unbound fraction of a drug at any time after a single intravenous dose is applied.

An important advantage of this method of calculating $c_{free}$ and $f_{uP}$ *in vivo* is satisfying Goldberg Waage law at any point of time after the drug application when quantitative determination of the total drug concentration is feasible.

The absence of the elimination rate constant in the formula is not an error because the parameter is hidden in the values of the total drug concentrations.

It is essential that the association constant ($K_a$) determined *in vitro* and the total amount of the drug ($\Lambda$) and the total drug concentration ($c_{total}$) are all obtained from the same serum.

This means that described procedure could be used in the phase of gathering information about the pharmacokinetic properties of the drug in both preclinical studies and clinical trials in phase I and II.

It seems that our observations provide an important complement to existing methods of analysis to pharmacokinetics processes.

We expect to verify the proposed model in further research on experimental data.

# REFERENCES

1.Toutain P. I., Bousquet-Melou A. (2002) Free drug fraction vs. free drug concentration: a matter of frequent confusion. J. Vet. Pharmacol. Therap. 25, 460–463.

2.Schmidt S., Gonzalez D., Derendorf H. (2010) Significance of protein binding in pharmacokinetics and pharmacodynamics Journal of Pharmaceutical Sciences Volume 99, Issue 3, pages 1107–1122, March 2010.

3.Smith D. A., Li Di., Kerns E. H. (2010) The effect of plasma protein binding on in vivo efficacy: misconceptions in drug discovery Nature Reviews Drug Discovery 9, 929-939 (December 2010) | doi:10.1038/nrd3287).

4.Waage, P. Guldberg, C. M. (1834) Forhandlinger: *Videnskabs-Selskabet i Christiana*, 35. expression (8).